\documentclass[twocolumn,showpacs,preprintnumbers,amsmath,amssymb,prb,floatfix,superscriptaddress]{revtex4}

\usepackage{graphicx,tabularx}
\usepackage{amssymb}
\usepackage{dcolumn}
\usepackage[mathcal]{euscript}
\usepackage{color}
\usepackage{multirow}
\usepackage[version=3]{mhchem} 
\usepackage{natbib}
\definecolor{gray0}{gray}{0.0}
\definecolor{gray64}{gray}{0.25}
\definecolor{gray128}{gray}{0.5}
\definecolor{gray192}{gray}{0.75}
\definecolor{gray255}{gray}{1.0}

\begin{document}
\title[Extending Hirshfeld-I Reply]{Reply to `Comment on ``Extending Hirshfeld-I to bulk and periodic materials'' '}
\author{D. E. P. Vanpoucke}
\affiliation{SCRiPTS group, Dept. Inorganic and Physical Chemistry, Ghent University, Krijgslaan $281$ - S$3$, $9000$ Gent, Belgium}
\affiliation{Ghent Quantum Chemistry Group, Dept. Inorganic and Physical Chemistry, Ghent University, Krijgslaan $281$ - S$3$, $9000$ Gent, Belgium}
\author{I. Van Driessche}
\affiliation{SCRiPTS group, Dept. Inorganic and Physical Chemistry, Ghent University, Krijgslaan $281$ - S$3$, $9000$ Gent, Belgium}
\author{P. Bultinck}
\affiliation{Ghent Quantum Chemistry Group, Dept. Inorganic and Physical Chemistry, Ghent University, Krijgslaan $281$ - S$3$, $9000$ Gent, Belgium}
\date{\today}
\begin{abstract}
The issues raised in the comment by T.A. Manz are addressed through the presentation of calculated atomic charges for \ce{NaF}, \ce{NaCl}, \ce{MgO}, \ce{SrTiO3} and \ce{La2Ce2O7}, using our previously presented  method for calculating Hirshfeld-I charges in Solids [J. Comput. Chem.. doi: 10.1002/jcc.23088]. It is shown that the use of pseudo-valence charges is sufficient to retrieve the full all-electron Hirshfeld-I charges to good accuracy. Furthermore, we present timing results of different systems, containing up to over $200$ atoms, underlining the relatively low cost for large systems. A number of theoretical issues is formulated, pointing out mainly that care must be taken when deriving new atoms in molecules methods based on ``expectations'' for atomic charges.
\end{abstract}

\pacs{  } 
\maketitle
\indent Our recent extension of the Hirshfeld-I method to solids and periodic systems \cite{VanpouckeDannyEP:JCC2012a}, allows for the calculation of atoms in molecules (AIM) in solid state codes using common techniques as pseudo-potentials and plane wave basis sets. As a Hirshfeld-type method,\cite{HirshfeldFL:1977TCA} it relies on atomic reference densities that are used to construct AIM weight functions $w(\mathbf{r})$ that allow to extract an AIM density function from a computed system electron density distribution (EDD). As was shown by us, when using pseudo-potentials and plane waves, some issues appear that require attention when generating atomic reference densities.\cite{VanpouckeDannyEP:JCC2012a} However, we showed that the delocalization problem can be handled, and that it is actually part of the larger conceptual problem of defining reference densities for anions. In addition, we showed that the use of pseudo-potentials, and their resulting pseudo-valence charges, can provide the all-electron values for the atomic charges. Finally, we showed that the method scales favorably for large systems.\\
\indent  In a comment to our paper,\cite{ManzTA:JComputChemReply2012} T. A. Manz raises some questions which we address here in detail. His criticisms and suggestions can be summarized as:
\begin{enumerate}
  \item Hirshfeld-I atomic charges do not give ``chemically feasible'' atomic charges and improved performance of Atoms in Molecules (AIM) methods may be obtained by combining spherical averaging and Hirshfeld-I methods.\cite{ManzTA:JComputChemReply2012}
  \item More (complex) solids should have been investigated.
  \item The presented computational scaling was insufficiently justified.
\end{enumerate}
\indent In the following paragraphs we address each of these comments, from a theoretical as well as a computational perspective. \par

\begin{table*}[tb!]
\caption{Hirshfeld-I (HI) atomic charges according to different models for the atomic reference density: the R1 and R3 reference atoms for pseudo-valence densities (psv), and R3 reference atoms for all-electron valence densities (aev) and total (including core) all-electron densities (ae).\cite{VanpouckeDannyEP:JCC2012a, fn:valence} Bader charges calculated using the Henkelman algorithm\cite{HenkelmanG:ComputMaterSci2006} of $2006$ are also presented in comparison.\cite{BaderRFW:book1990, BaderRFW:1991ChemRev}}
\label{table:AIMmodelcharges}
\begin{ruledtabular}
\begin{tabular}{l|rrrrrrrrr}
& \ce{NaF} & \ce{NaCl} & \ce{MgO} & \multicolumn{3}{c}{\ce{SrTiO3}} & \multicolumn{3}{c}{\ce{La2Ce2O7}} \\
& q$_{\mathrm{Na}}$ & q$_{\mathrm{Na}}$ & q$_{\mathrm{Mg}}$ & q$_{\mathrm{Sr}}$  & q$_{\mathrm{Ti}}$ & q$_{\mathrm{O}}$ & q$_{\mathrm{La}}$ & q$_{\mathrm{Ce}}$ & q$_{\mathrm{O}}$ \\
\hline\\[-2mm]
Bader    & $0.85$ & $0.85$ & $1.67$ & $1.55$ & $2.17$ & $-1.24$ & $1.88/2.12$ & $2.33/2.38$ & $-1.21/-1.28$ \\
HI R1 psv& $1.04$ & $1.04$ & $1.31$ & $1.28$ & $2.57$ & $-1.29$ & $2.17/2.33$ & $2.64/2.79$ & $-1.41/-1.43$ \\
HI R3 psv& $1.04$ & $1.04$ & $1.57$ & $1.65$ & $2.67$ & $-1.44$ & $2.32/2.47$ & $2.81/2.94$ & $-1.49/-1.52$ \\
HI R3 aev& $1.04$ & $1.04$ & $1.56$ & $1.64$ & $2.66$ & $-1.43$ & $2.29/2.44$ & $2.78/2.91$ & $-1.49/-1.52$ \\
HI R3 ae & $1.05$ & $1.05$ & $1.62$ & $1.62$ & $2.69$ & $-1.43$ & $2.07/2.21$ & $2.46/2.60$ & $-1.43/-1.47$ \\
\end{tabular}
\end{ruledtabular}
\end{table*}

\paragraph{Atoms in molecules and atomic charges\\}
\indent Atomic charges are without doubt useful quantities to understand molecular properties and even, to some extent, make predictions. However, there are a number of critical issues listed below that make comparisons difficult.
\begin{itemize}
  \item AIM are not observables and hence the classical-quantum correspondence principle cannot be applied to lead to a unique operator that, acting on a wave function, gives the AIM. This explains the wealth of methods that has been introduced.\cite{BultinckPopelier:book2009} The fact that no observable can be associated with atomic charges also entails that criticisms on AIM charges are hard to substantiate. Yet, quite fierce discussions have appeared, most prominently in the context of whether an AIM can be uniquely defined \cite{ParrRGAyersPWNalewajskiRF:2005JPCA, BaderRFWMattaCF:2004JPhysChemA, BaderRFWMattaCF:2006JPhysChemA} and in discussions of whether Bader charges are too large.\cite{GuerraCF:JComputChem2004, BaderRFWMattaCF:2006JPhysChemA} As for the unique definition, we share the point of view that no unique method can exist because there simply is no beacon that allows us to uniquely and unequivocally define AIM\cite{ParrRGAyersPWNalewajskiRF:2005JPCA} and thus establish what charges are ``correct''.
  \item Manz speaks of ``chemically feasible'' or ``more realistic'' atomic charges without reference to what physical law or theorem or similar has been used to establish what is ``chemically feasible'' or ``more realistic''. We can only guess that either some reasoning is applied based on oxidation numbers (in reference \onlinecite{ManzSholl:JChemTheoryComp2010} for \ce{SrTiO3} Manz and Sholl state that ``one expects the NACs to lie between zero and the oxidation states of $+2$ (Sr), $+4$ (Ti), and $-2$ (O)'') or that there is some general assumption that, at least for positively charged AIM, the positive charge always has to be between zero and maximally the number of valence electrons (``\emph{Results for bulk \ce{SrTiO3} provide a useful example. Because \ce{Sr} has two valence electrons, its net atomic charge should be less than or equal to two.}'' taken from  Ref.~\onlinecite{ManzTA:JComputChemReply2012}). Old school chemical reasoning uses a series of useful, yet very qualitative, rules; such as attaching a charge of $+2$ to alkaline earth metal ions, $-2$ to oxygen atoms etc. (with known exceptions, obviously). We strongly believe that one should abandon such reasoning and not trust these as strict guidelines. Whether the charge should be $1.54$ or $2.50$, as was calculated in the preceding comment,\cite{ManzTA:JComputChemReply2012} using the Bader method and the Hirshfeld-I method, respectively, is a question that cannot be answered. Classical reasoning based on ionization energies and electron affinities, or electronegativities of free atoms, has its limitations and once an atom is embedded in a molecule, the free atomic properties loose their physical meaning. Charge transfer will take place until the electronegativities of all AIM are exactly the same and exactly the same as the molecular electronegativity.\cite{SandersonRT:Science1951, SandersonRT:JACS1983, ParrRGYangW:1989DFT_AandM}
  \item The Hirshfeld-I method was introduced to improve the Hirshfeld method, not so much because the charges were \textit{chemically} too small but because they were too small due to \textit{mathematical} issues related to the information measure used.\cite{BultinckP:2007JCP_HirRef} AIM methods can, however, be substantiated by physical principles and computational properties. Although admittedly mostly established for molecular quantum chemical calculations until now, it is reassuring that Hirshfeld-I is  among the least basis set dependent methods \cite{BultinckPAyersPWFiasS:2007ChemPhysLett} and is observed to always yield the same result.\cite{BultinckPAyersPWFiasS:2007ChemPhysLett, GhillemijnD:JCompuChem2011} From a theoretical perspective, we choose to follow the path of establishing an AIM method from elegant and simple principles without need to rely on combining AIM methods using some weighting\cite{ManzSholl:JChemTheoryComp2010} or introducing some fitting parameters (such as some ``carefully chosen radius''\cite{ManzTA:JComputChemReply2012}). The Hirshfeld-I method is further backed up by the fact that it gives good quality electrostatic potentials\cite{VanDamme_Bultinck:JCTC2009, VerstraelenT:JPhysChemC2012} and performs well in Electronegativity Equalization Methods (EEM)\cite{MortierWJ:JACS1986, ParrRGYangW:1989DFT_AandM}. This requires as good as possible transferability of the AIM and it is reassuring that Hirshfeld-I performs among the very best methods \cite{VerstraelenT:JChemPhys2009,VerstraelenT:JChemTheorComput2011}. The Hirshfeld method, for example, is not compatible with EEM models\cite{BultinckP:JPhysChemA2002} and in our experience, ISA also does not perform well in EEM. ISA does perform better at reproducing the ESP, which is well expected given that it uses more degrees of freedom (one per radius where the spherical averaging is taken) than Hirshfeld-I. On the other hand, the fact that it does not perform well in EEM is due to the poorer transferability. This agrees with earlier findings of Manz and Sholl \cite{ManzSholl:JChemTheoryComp2010}.
  \item In the work of Manz and Sholl,\cite{ManzSholl:JChemTheoryComp2010} a setup is used that gives what Manz considers ``improved performance''\cite{ManzTA:JComputChemReply2012} or ``more realistic'' charges.  The solution of Manz and Sholl to combine the iterative stockholder approach (ISA)\cite{LillestolenTCWheatleyRJ:2008ChemComm, LillestolenTCWheatleyRJ:2009JChemPhys} and Hirshfeld-I may be a good pragmatic choice for their needs but is not desirable given the fact that this combines the weak point of both methods in one single method. As described below; Hirshfeld-I has issues with anionic reference species, such that special precautions have to be taken. As described by one of us in $2009$ for systems with a fairly dense or spherical coordination around a specific atom,\cite{BultinckP:PhysChemChemPhys2009} the ISA AIM will artificially allocate density from far away. This was confirmed in 2010 by Manz and Sholl \cite{ManzSholl:JChemTheoryComp2010} for crowded systems. The two may balance in the application by Manz but such compensations are neither reliable, nor elegant. As an example: an endohedral coordination of Li in \ce{C60} with a global charge $+1$ results in an ISA charge for \ce{Li} of $-5.895$.\cite{fn:privCommToon} Having to implement checks whether both methods are applicable to a specific problem at hand is undesirable. In addition, having to introduce constraints may introduce a degree of arbitrariness and renders the procedure less elegant. However ingenious, the DDEC/c3 method needs to rely on conditioning of reference densities, constraints on decays in densities and on the number of valence electrons on every atom to keep these positive\cite{ManzTAShollDS:JChemTheorComput2012}. We follow the Hirshfeld-I path which combines optimally the transferability of the AIM and their performance for electrostatic potentials, even across conformations\cite{VerstraelenT:JChemTheoryComput2012}, a fact confirmed by Manz and Sholl\cite{ManzTAShollDS:JChemTheorComput2012}.
  \item Due to the fact that no definite correct AIM model and thus atomic charges are known, the only possible test cases for a computer program are precisely those where due to symmetry some charges have to be zero, and thus the program should give zero charges. So rather than considering diamond and graphene to be application tests, they are validation tests. Hence, the fact that a zero charge is indeed found, validates not only the method but also the code. Bugs in a computer code cannot be identified from other applications as the ``right answer'' is not known. This is the reason why such tests have been applied.\cite{VanpouckeDannyEP:JCC2012a} However, as will be shown below, our Hirshfeld-I implementation works well for several of the systems suggested as tests by Manz.\cite{ManzTA:JComputChemReply2012}
  \item We disagree with the claim by Manz\cite{ManzTA:JComputChemReply2012} that Hirshfeld-I fails to give ``chemically feasible'' charges. It may ``malfunction" in terms of his requirements for an AIM method \textit{when} using his implementation, convergence criteria and reference densities but as we show below: in our case it always gives results that would be considered ``chemically feasible''.
\end{itemize}
\indent Hirshfeld-I does have its issues, as any method, and we agree fully with Manz and Sholl\cite{ManzTAShollDS:JChemTheorComput2012} that \emph{the} main problem are the anionic reference densities.\cite{ManzTA:JComputChemReply2012} The approach taken by Manz and Sholl\cite{ManzSholl:JChemTheoryComp2010} using background charges is definitely one that we will pursue, next to many other paths being explored, including Watson spheres.\cite{WatsonRE:PhysRev1958} On the other hand, the R3 and R4 methods already succeed at eliminating this problem. Moreover, as we show below, using our method that also allows avoiding the rise in electron density for \ce{O^{2-}} at large distance, the AIM charges do not change significantly when using either valence-only densities or the full all-electron densities. The charges for \ce{Sr} stay in all cases nicely below $2.00$, a feature sought by Manz. As a final remark, information loss measures do not establish that one should always consider the entire electron density. We do agree that using the entire density is the most attractive but it can be done for valence only, or core and valence separately,\cite{VanpouckeDannyEP:JCC2012a}. We do iterate again, that using the setup published by us,\cite{VanpouckeDannyEP:JCC2012a} the AIM charges are the same whether we use valence densities only or all-electron densities.\par

\begin{figure*}
  \includegraphics[width=14.0cm,keepaspectratio=true]{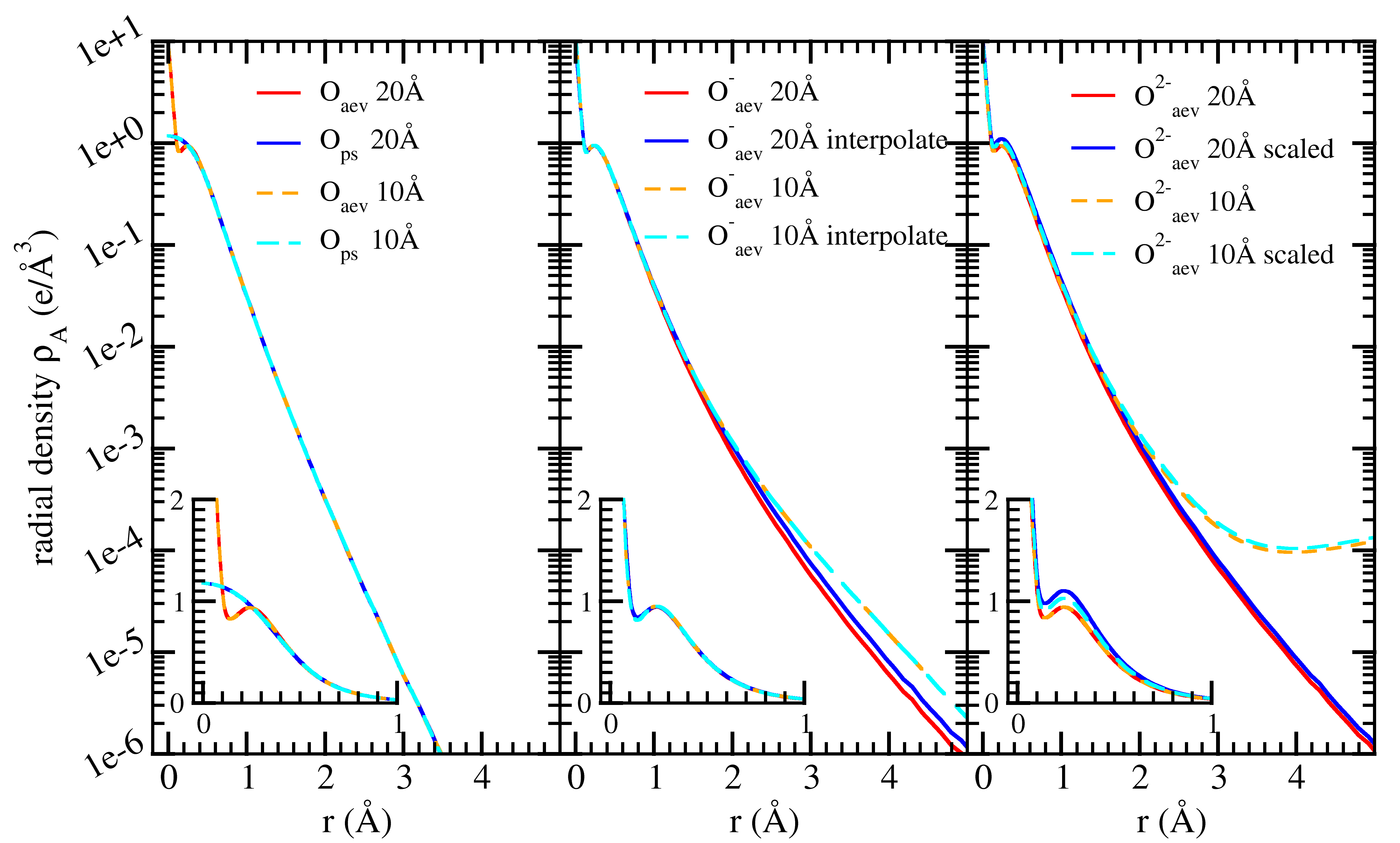}\\
  \caption{Radial EDDs for different O reference ions.}\label{fig:Oradii}
\end{figure*}

\paragraph{Computed atomic charges in ionic compounds\\}
\indent We have calculated the atomic charges for a set of systems considered by Manz\cite{ManzTA:JComputChemReply2012} (\ce{NaF}, \ce{NaCl}, \ce{MgO}, and \ce{SrTiO3}) using our previously presented method.\cite{VanpouckeDannyEP:JCC2012a} All system EDDs are obtained within the DFT framework with LDA functionals, and using sufficiently large $\Gamma$-centered $21\times 21\times 21$ k-point sets (for computational details see Vanpoucke et al.\cite{VanpouckeDannyEP:JCC2012a}). In addition, atomic charges for pyrochlore structure \ce{La2Ce2O7} are presented, using the system EDD obtained during our previous study.\cite{VanpouckeDannyEP:PhysRevB2011} Hirshfeld-I charges are calculated using radial atomic EDDs obtained using the R1 and R3 methods, presented previously.\cite{VanpouckeDannyEP:JCC2012a} Valence only reference atoms obtained from pseudo-densities are labeled as ``psv'', while valence only reference atoms obtained from all-electron calculations are labeled as ``aev''.\cite{fn:valence} The used full all-electron reference atomic EDDs are the sum of all-electron valence densities and all-electron core densities and are labeled ``ae''. For all these systems, Lebedev grids of $1202$ grid points per spherical shell are used, and the convergence criterion on the atomic charges is set to $< 1.0\times 10^{-5}\ e$.\\
\indent As commented above, we suspect that in the view of Manz \cite{ManzTA:JComputChemReply2012}, atomic charges are considered ``chemically feasible'' for an atom with positive AIM charge if its charge is smaller than the number of valence electrons or if the magnitude of the charge is below the oxidation state \cite{ManzSholl:JChemTheoryComp2010}. As several of the systems in Table~\ref{table:AIMmodelcharges} were also studied by Manz,\cite{ManzTA:JComputChemReply2012} a comparison can be made. Taking as an example \ce{MgO}, we find that the atomic charges reported by Manz are roughly $2.03$, corresponding to complete ionicity. The degree of ionicity is much smaller in the case of Bader AIM\cite{BaderRFW:1991ChemRev, BaderRFW:book1990} (note, our Bader data are close to, but do not fully coincide with the value of Manz (differences around $0.05$). This may be due to shortcomings in the $2006$ implementation by Henkelman \textit{et al.} \cite{HenkelmanG:ComputMaterSci2006, TangW:JPhysCondensMatter2009} or minor differences in technicalities of the calculations.) All our reported Hirshfeld-I R3 results are relatively closer to the Bader results than to the value of $2.03$ obtained using the DDEC methods. The key point is that all point in the same direction for the charge transfer, but the degree to which this happens differs. As atomic charges are not an observable, it is hard to decide which is ``correct''. In this context, it is worth noting that also from an experimental point of view, different analyses of XRD data give different results; which in some cases points toward perfect ionicity or in other cases to \ce{Mg+} (\textit{cf.} Refs.~\onlinecite{VidalValatG:ActaCrystA1978, ZuoJM:PhysRevLett1997, TsirelsonVG:ActaCryst1998, IsraelS:JPhysChemSolids2003} for a sample of the theoretical and experimental data on this, at first glance, simple system). In any event, if ``chemical feasibility'' is defined as we suspect, all \ce{Mg} charges are still ``chemically feasible''.\\
\indent As is shown in Table~\ref{table:AIMmodelcharges}, the R1 and R3 reference sets yield different atomic charges. This was already noted in our original paper,\cite{VanpouckeDannyEP:JCC2012a} and identified to be a consequence of the fact that the integrated reference densities for the anions lack some density. By correcting this charge discrepancy (R3 and R4 reference densities), the obtained results show better agreement with molecular quantum mechanical calculations for the investigated set of molecules.\cite{VanpouckeDannyEP:JCC2012a, VanDamme_Bultinck:JCTC2009}\\
\indent Comparison of the charges obtained using psv, aev, or ae system and atomic reference EDDs shows that all three give essentially the same results. Note that the larger gradients present in the all-electron calculations require finer meshes for the density grids of the systems under study. This becomes quite clear for heavy atoms such as \ce{La} and \ce{Ce} in the \ce{La2Ce2O7} system. This clearly shows the advantage of our approach, by being able to obtain all-electron quality charges using valence only densities,\cite{fn:valence} one can use much coarser grids to store the system EDDs. In addition, our method does not need to make use of all-electron core densities obtained from other sources (\textit{e.g.} molecular quantum chemical calculations), allowing us to refrain from mixing the results of different approaches.\\
\indent Comparison to the Hirshfeld-I charges reported in the preceding comment\cite{ManzTA:JComputChemReply2012} shows there is a significant discrepancy for \ce{MgO} and \ce{SrTiO3}. Note that our values presented for these two materials seem ``chemically feasible'' according to Manz. As was shown in our paper,\cite{VanpouckeDannyEP:JCC2012a} the delocalization of electrons to infinity for anions, makes it quite difficult to generate qualitatively good atomic reference EDDs (this in addition to the conceptual issue of defining reference densities for anions). It was shown that for anions, even unit cells of $20\times 20\times 20$\AA$^3$ are too small to obtain good EDDs for anions.\cite{VanpouckeDannyEP:JCC2012a} In addition to tail effects, the delocalization of electrons also results in electron deficiency for almost all anions upon integration of the radial EDD. In case of the DDEC code, the reference atoms in the c1 method are obtained in a unit cell as small as $10\times 10\times 10$\AA$^3$.\cite{ManzSholl:JChemTheoryComp2010} In Fig.~\ref{fig:Oradii} we show the ionic radial EDD for \ce{O}, \ce{O-}, and \ce{O^{2-}}. Although a $10\times 10\times 10$\AA$^3$ unit cell gives the same results for the neutral atom as a $20\times 20\times 20$\AA$^3$ unit cell, it is clear that for the anions, the radial EDD in the small unit cell becomes much larger (too large) already at distances $<3$\AA. For the \ce{O^{2-}} there is a clear contribution due to overlap of the EDD of the periodic copies, in addition to delocalized electrons at the center of the unit cell, when the $10\times 10\times 10$\AA$^3$ unit cell is used. This delocalization of electrons, results in a lack of electrons upon integration of the radial EDD of the reference anions (although part of this lack is compensated due to the overlap between periodic copies). The use of larger unit cells is an essential part for generating qualitatively good atomic reference densities, and we found the R3 and R4 methods to be able to resolve the problem of electron deficiency due to electron delocalization.\cite{VanpouckeDannyEP:JCC2012a} So we stress that with our R3 and R4 methods all Hirshfeld-I charges are ``chemically feasible''.\\

\begin{figure}[t!]
  \includegraphics[width=8cm,keepaspectratio=true]{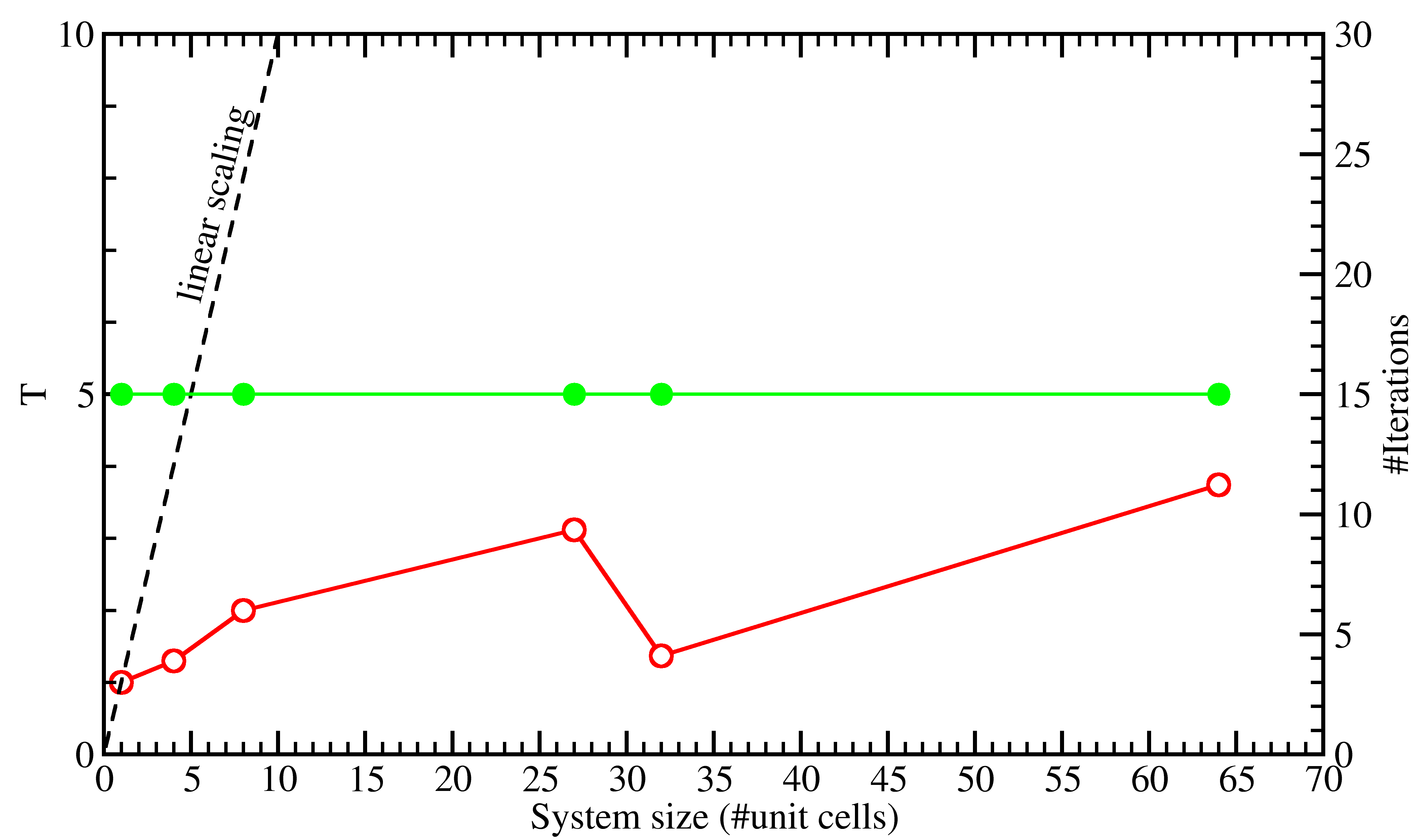}\\
  \caption{Scaling behavior of the implemented method for \ce{NaF}. The red circles give the relative CPU time T, as defined in Eq.\eqref{eq:Tdef} for the different supercells, while the green discs show the number of iterations required for a Hirshfeld-I calculation for each of the supercells. Note that the system size is defined as the number of unit cells, not atoms. To obtain the number of atoms in the \ce{NaF} supercells, the system size shown needs to be multiplied by $2$. The dashed line indicates linear scaling.}\label{fig:scaling}
\end{figure}

\paragraph{Computational cost\\}
\indent The computational scaling of an AIM method as the one presented is important as it determines the extent to which the method will be used. However, such information is not always sufficiently well presented. According to the preceding comment,\cite{ManzTA:JComputChemReply2012} our scaling example using diamond supercells insufficiently clearly showed on the one hand how our Hirshfeld-I calculations scale and on the other hand that for systems containing many inequivalent atoms such calculations are still easily feasible. To obtain clear scaling information with regard to system size for a system with inequivalent atoms, we have used the \ce{NaF} system and performed Hirshfeld-I calculations for different supercell sizes.\\
\indent Because the absolute values for the CPU time depend on the CPU architecture and the number of cores used, we performed all calculations on the same machine, using the same number of cores. In the following, we present the relative CPU time T for a Hirshfeld-I calculation on a system in reference to the CPU time required for a Hirshfeld-I calculation on the \ce{NaF} unit cell:
\begin{equation}\label{eq:Tdef}
\mathrm{T}=\frac{\mathrm{CPU}(X)}{\mathrm{CPU}(\mathrm{NaF}\ 1)} .
\end{equation}
\indent Table~\ref{table:scaling} shows the size of the different systems using different measures. For practical applications, the number of atoms is the only size known before the start of a Hirshfeld(-I) calculation, and thus most of interest for actual applications. In addition, we also give the size of the SoI, the total number of grid points in the spherical integration grids (Sph.gr.), and the number of iterations (iter.) required to obtain convergence. In all cases the same convergence criterion of $1\times 10^{-5}\ e$ was used. An integration precision, defined as the percentage of the electrons accounted for by the AIM populations, of at least $99.98$\% was obtained for all systems. A Lebedev grid of $1202$ grid points per spherical shell was used.\cite{LebedevVI_grid:1999DokladyMath}\\
\indent The \ce{NaF} series, consisting of \ce{NaF} calculations using different supercells, shows that the size of the supercell used has no influence on the number of iterations (\textit{cf.} Fig.~\ref{fig:scaling}), as was already noted earlier for the diamond series.\cite{VanpouckeDannyEP:JCC2012a}\\
\indent Figure~\ref{fig:scaling} also shows the scaling behavior for the \ce{NaF} system as function of the number of unit cells in the supercell. As for the diamond system, a clearly sublinear  scaling is observed. The dip in the \ce{NaF} timing curve may at first seem a bit odd. This, however, is due to the shape of the supercells. Since \ce{NaF} is face centered cubic, primitive supercells were used for all sizes except the second and fifth. These were constructed using a cubic supercell (similar to our previous work on cerates\cite{VanpouckeDannyEP:PhysRevB2011, VanpouckeDannyEP:ApplSurfSci2012}). Because the cubic supercell is more compact, it gives rise to a smaller SoI, which in turn results in a smaller integration grid.\\

\indent In addition, we present timings for a set of different periodic  systems containing a varying number of inequivalent atoms. These are the \ce{NaCl}, \ce{MgO}, and \ce{SrTiO3} unit cells, cubic \ce{CeO2} supercells doped with $3.125$\% group IV elements, a special quasi random structure (SQS) $88$ atom supercell of \ce{La2Ce2O7}, and Pt induced Ge nanowires on a Ge(001) surface of types NW1 and NW2, the latter also with a CO molecule adsorbed on a B3 site (NW2+CO). Since here we are only interested in timings, we do not discuss the obtained charges for these systems, this will be done elsewhere (\textit{e.g.} Ref.~\onlinecite{VanpouckeDannyEP:XX2012}). A more detailed description of the SQS system, and the computational setup used during the ab initio calculation of its properties was previously presented.\cite{VanpouckeDannyEP:PhysRevB2011} For the issue of interest it is important to know that such a system is constructed to mimic a crystal lattice with a truly random distribution of two or more types of ions.\cite{Zunger_SQS:PRB1990, Zunger_SQS:PRL1990} In this sense, all atoms in a SQS can be considered inequivalent. The different nanowire structures and their CO adsorption sites were previously investigated by one of us,\cite{vanpoucke:prb2008R, Vanpoucke:mrs09eprocNW, Vanpoucke:prb2010NW, Vanpoucke:PRB2010COonNW} and the geometries for the systems used are shown in Fig. $12$a (NW1) and $15$a (NW2) of Ref.~\onlinecite{Vanpoucke:prb2010NW}, while the B3 adsorption site for CO is shown in Fig.~$1$b of Ref.~\onlinecite{Vanpoucke:PRB2010COonNW}. Since the nanowire systems are surface systems, a large part of the unit cell consists of a vacuum region separating periodic copies of the surface slab. This reduces the Sphere of Influence (SoI) somewhat.\cite{VanpouckeDannyEP:JCC2012a} Surface reconstructions on the other hand, result in a large number of inequivalent atoms in the system.\\
\indent The series of doped \ce{CeO2} systems, shows that quite similar systems (only one atom difference) in this case also require a similar number of iterations. The nanowire systems, however, show this is not always the case; and that for example the adsorption of a molecule can influence the number of required iterations significantly. However, from Table~\ref{table:scaling} no immediate relationship between the number of Hirshfeld-I iterations on the one hand and system size, SoI or the spherical integration grid, on the other hand, appears to be present.\\

\begin{table}[tb!]
\caption{Size information for the set of studied systems (\textit{cf.} text): system size (\#atoms), sphere of influence (SoI), spherical integration grid (Sph.gr), and number of iterations (iter.). In addition, the relative CPU time T, as defined in Eq.\eqref{eq:Tdef}, is also presented.}
\label{table:scaling}
\begin{ruledtabular}
\begin{tabular}{l|rrrrr}
system & \#atoms & SoI & Sph.gr. & iter. & T\\
       &        &     & ($\times 10^6$) & \\
\hline
\ce{NaF} 1 & $2$   & $3378$ & $19$ & $15$ & $1.0$\\
\ce{NaF} 2 & $8$   & $3720$ & $24$ & $15$ & $1.3$\\
\ce{NaF} 3 & $16$  & $4598$ & $34$ & $15$ & $2.0$\\
\ce{NaF} 4 & $54$  & $6022$ & $53$ & $15$ & $3.1$\\
\ce{NaF} 5 & $64$  & $5600$ & $49$ & $15$ & $1.4$\\
\ce{NaF} 6 & $128$ & $7662$ & $77$ & $15$ & $3.7$\\
\hline
\ce{NaCl}  & $2$ & $1942$ & $12$ & $17$ & $0.4$\\
\ce{MgO}   & $2$ & $4242$ & $24$ & $58$ & $6.6$\\
\ce{SrTiO3}& $5$ & $3510$ & $23$ & $55$ & $4.3$\\
\ce{La2Ce2O7} SQS & $88$ & $5714$ & $64$ & $27$ & $3.1$ \\
\hline
\ce{CeO2}+\ce{C}  & $96$ & $5916$ & $56$ & $21$ & $1.4$\\
\ce{CeO2}+\ce{Si} & $96$ & $5976$ & $56$ & $26$ & $1.6$\\
\ce{CeO2}+\ce{Ge} & $96$ & $5964$ & $56$ & $22$ & $1.4$\\
\ce{CeO2}+\ce{Sn} & $96$ & $5916$ & $56$ & $25$ & $1.6$\\
\ce{CeO2}+\ce{Pb} & $96$ & $5916$ & $56$ & $22$ & $1.4$\\
\ce{CeO2}+\ce{Ti} & $96$ & $5976$ & $56$ & $21$ & $1.4$\\
\ce{CeO2}+\ce{Zr} & $96$ & $5916$ & $56$ & $23$ & $1.5$\\
\ce{CeO2}+\ce{Hf} & $96$ & $5916$ & $63$ & $22$ & $1.5$\\
\hline
NW1         & $100$ & $1904$ & $30$ & $55$ & $1.7$\\
NW2         & $202$ & $2375$ & $42$ & $83$ & $7.9$\\
NW2+\ce{CO} & $206$ & $3019$ & $47$ & $108$& $12.4$ \\
\end{tabular}
\end{ruledtabular}
\end{table}

\indent Table~\ref{table:scaling} shows the timing results for the different systems, as a function of the system size. The size varies over two orders of magnitude, going from the very small $2$ atom unit cells up to the $206$ atom large nanowire system with adsorbed CO molecules.\cite{vanpoucke:prb2008R, Vanpoucke:mrs09eprocNW, Vanpoucke:prb2010NW, Vanpoucke:PRB2010COonNW, VanpouckeDannyEP:PhysRevB2011, VanpouckeDannyEP:ApplSurfSci2012} The ab initio calculations required to obtain the system EDDs for the former can easily be run on any desktop machine, while the latter require a sizeable supercomputer. However, as can be seen in Fig.~\ref{fig:scaling}, the required CPU time for Hirshfeld-I calculations on both types of systems scale in a much better way. This is mainly due to the comparable size of the SoI, and allows us to reiterate our statement that our implementation is able to easily handle much larger systems, and that the limiting factor is the required ab initio calculation of the system EDD.\\
\indent Table~\ref{table:scaling} also shows that the computational cost is not linked to the complexity of the system (\textit{i.e.} the number of inequivalent atoms). If this were the case, the SQS system would have to be one of the most expensive systems, which it is not. With convergence in only $27$ iteration steps, it is comparable to the presented cerates.\\

\indent In conclusion, because atomic charges are not observables, and as such, absolute values are not available, we opt to introduce atomic charges as results of an algorithm based on some clear mathematical or physical-chemical reasoning and insist on using methods that perform well in electronegativity equalization, a solid physical theorem. Admittedly, anionic densities form a weak point in the entire Hirshfeld-I setup although in the present setup for solid state calculations this problem has been alleviated.\\
\indent The results presented in Table~\ref{table:AIMmodelcharges}, using our implementation, show that for more complex systems:
\begin{itemize}
  \item ``chemically feasible'' charges are obtained,
  \item valence only EDDs are sufficient to obtain full all-electron charges, and this using a much coarser grid for the system EDD,
\end{itemize}
\indent With regard to the computational cost, we have shown that, for a set of quite different systems:
\begin{itemize}
    \item the number of iterations does not depend on the system size, and does not change when using a different size supercell for a given material,
    \item the increase of the computational cost with regard to the system size is limited, making the study of larger systems easily feasible.
\end{itemize}

\section*{Acknowledgement}
\indent The research was financially supported by FWO-Vlaanderen, project n$^{\circ}$ $3$G$080209$. This work was carried out using the Stevin Supercomputer Infrastructure at Ghent University, funded by Ghent University, the Hercules Foundation and the Flemish Government--department EWI.



\begin{thebibliography}{47}
\bibitem[{fn:()}]{fn:valence}
\bibinfo{note}{Note that in this work and in our previous work, we refer to
  valence electrons as the electrons that are or may be involved in the
  chemistry of the system.\cite{VanpouckeDannyEP:JCC2012a} This means that
  filled shells close to the Fermi level are also included. For example, the
  $3d$ shell of \ce{Ge}, and the $5s$ and $5p$ shells of \ce{Ce} are included.}
\end{thebibliography}

\end{document}